\documentclass[
    aps,
    pra,
    superscriptaddress,
    twocolumn,
    a4paper,
    floatfix
]{revtex4-2}

\usepackage[american]{babel}
\usepackage[utf8x]{inputenc}

\usepackage{grffile} 

\usepackage{newtxtext,newtxmath}
\usepackage{microtype}

\usepackage{amsmath}
\usepackage{amssymb}
\usepackage{bbm}
\usepackage{mathtools}
\usepackage{dsfont}
\usepackage{braket}
\usepackage{cancel}
\usepackage{slashed}

\usepackage{graphicx}
\usepackage[svgnames]{xcolor}

\usepackage{siunitx} 

\usepackage{ragged2e}
\usepackage{array}
\usepackage{tabularx}
\usepackage{booktabs}

\definecolor{cset-aps-blueberry}{RGB}{28,128,158}
\definecolor{cset-aps-blue}{RGB}{46,44,184}
\definecolor{cset-aps-turquoise}{RGB}{0,67,88}
\definecolor{cset-aps-limegreen}{RGB}{190,219,67}
\definecolor{cset-aps-green}{RGB}{31,138,112}
\definecolor{cset-aps-yellow}{RGB}{255,225,25}
\definecolor{cset-aps-orange}{RGB}{253,116,0}
\definecolor{cset-aps-red}{RGB}{219,0,43}

\usepackage{tikz}

\usepackage{pgfplots}
\pgfplotsset{%
    every axis legend/.append style={%
        cells={anchor=west},
        at={(0.96,0.04)},
        anchor=south east,
        font=\scriptsize,
        },
    every axis/.append style={%
        yticklabel style={%
            /pgf/number format/fixed zerofill,
            /pgf/number format/precision=2},
        },
    width= \textwidth,
    height=8cm,
    xmajorgrids=true,
    xminorgrids=false,
    minor x tick num=1,
    compat = 1.16
}
\usepgfplotslibrary{external}

\usetikzlibrary{decorations}

\usepackage{pict2e,picture}

\makeatletter
\DeclareRobustCommand{\Arrow}[1][]{%
\check@mathfonts
\if\relax\detokenize{#1}\relax
\settowidth{\dimen@}{$\m@th\rightarrow$}%
\else
\setlength{\dimen@}{#1}%
\fi
\sbox\z@{\usefont{U}{lasy}{m}{n}\symbol{41}}%
\begin{picture}(\dimen@,\ht\z@)
\roundcap
\put(\dimexpr\dimen@-.7\wd\z@,0){\usebox\z@}
\put(0,\fontdimen22\textfont2){\line(1,0){\dimen@}}
\end{picture}%
}
\makeatother

\usepackage{hyperref}
\hypersetup{%
    colorlinks=true,
    linkcolor={cset-aps-red},
    linkbordercolor={cset-aps-red},
    filecolor={cset-aps-orange},
    filebordercolor={cset-aps-orange},
    citecolor={cset-aps-blue},
    citebordercolor={cset-aps-blue},
    urlcolor={cset-aps-green},
    urlbordercolor={cset-aps-green},
    menucolor={cset-aps-limegreen},
    menubordercolor={cset-aps-limegreen},
    breaklinks=true,
    pdfborderstyle={/S/U/W 2},
    pdfpagemode=UseOutlines,
    pdfstartpage={1},
}

\newcommand{\ee}{\text{e}}
\newcommand{\ii}{\text{i}}

\usepackage{lipsum}
\usepackage{nicefrac}

\newcommand{\op}[1]{\hat{#1}}
\newcommand{\realpart}[1]{\mathrm{Re}\left({#1}\right)}
\newcommand{\imgpart}[1]{\mathrm{Im}\left({#1}\right)}

\newcommand{\ex}[1]{\exp\left(#1\right)}

\newcommand{\br}[1]{\Bra{#1}}		
\newcommand{\kt}[1]{\Ket{#1}}		
\newcommand{\kb}[2]{\Ket{#1}\Bra{#2}}		
\newcommand{\com}[2]{\left[#1,#2\right]}

\newcommand{\R}{\mathbb{R}}

\newcommand{\heavi}{\theta}
\newcommand{\definer}{:=}
\newcommand{\abs}[1]{\left| #1 \right|}
\newcommand{\tr}[1]{\mathrm{tr}\!\left\{#1\right\}}
\newcommand{\trin}[2]{\mathrm{tr}_{\mathrm{#1}}\!\left\{#2\right\}}
\newcommand{\mean}[1]{\Braket{#1}}
\def\dd{\ensuremath\mathrm{d}}

\begin{document}

\title{Qubit-based momentum measurement of a particle}
\collaboration{This article has been published in \href{https://link.springer.com/article/10.1140\%2Fepjd\%2Fs10053-021-00282-6}{The European Physical Journal D \textbf{75}, 271 (2021)};\\ 
licensed under a \href{http://creativecommons.org/licenses/by/4.0/}{Creative Commons Attribution [CC BY]} license.}
\author{Bernd Konrad} 
\email{bernd.konrad@uni-ulm.de}
\address{Institut für Quantenphysik and Center for Integrated Quantum Science and Technology ($\text{IQ}^{\text{ST}}$), Universität Ulm, Albert Einstein Allee 11, D-89069 Ulm, Germany}
\author{Fabio Di Pumpo}
\address{Institut für Quantenphysik and Center for Integrated Quantum Science and Technology ($\text{IQ}^{\text{ST}}$), Universität Ulm, Albert Einstein Allee 11, D-89069 Ulm, Germany}
\author{Matthias Freyberger}
\address{Institut für Quantenphysik and Center for Integrated Quantum Science and Technology ($\text{IQ}^{\text{ST}}$), Universität Ulm, Albert Einstein Allee 11, D-89069 Ulm, Germany}

\begin{abstract}
\noindent 
An early approach to include pointers representing measurement devices into quantum mechanics was given by von Neumann. 
Based on this idea, we model such pointers by qubits and couple them to a free particle, in analogy to a classical time-of-flight arrangement. 
The corresponding Heisenberg dynamics leads to pointer observables whose expectation values allow us to reconstruct the particle's momentum distribution via the characteristic function. 
We investigate different initial qubit states and find that such a reconstruction can be considerably simplified by initially entangled  pointers.
\end{abstract}

\maketitle

\section{Introduction}
Despite its rigorous axiomatic framework~\cite{VonNeumann1968MathematischeGrundlagen,gallone2014hilbert,david2014formalisms} 
and various conceptual studies~\cite{jammer1966conceptual,bohm1989quantumtheory,peres1995quantum,castro2019Introduction,Camilleri2009-CAMAHO}, many operational processes which are well-known in classical mechanics  are not intuitive in quantum mechanics~\cite{lamb1969operational,PhysRevD.24.1516,PhysRevD.26.1862,busch1997operational,klyshko1998basic,dariano2007Operational,zeh1970interpretation,zeh1973toward,brasil2015understanding}.
The formal description of a measurement requires the abstract projector formalism to gain knowledge about possible values for so-called observables.
This way, we can mathematically deduce the statistics of an observable associated with a specific state of the quantum system.

However, this process is not directly linked to an explicit experiment as a realization to measure this observable.
An early approach to remedy this problem was discussed by von Neumann in his classic textbook~\cite{VonNeumann1968MathematischeGrundlagen}. 
Pointer systems representing measurement devices were included into the description of the measurement process.
Hence, it is interesting to ask whether this idea allows us to assign basic classical concepts like time-of-flight (ToF) measurements an operational meaning in quantum mechanics~\cite{grot1996time,delgado1997arrival,leon_time--arrival_1997,muga2000arrival,anastopoulos_time--arrival_2006,anastopoulos_time--arrival_2012,anastopoulos_time--arrival_2017,anastopoulos_time_2019}.
In contrast to Ref.~\cite{grot1996time} we explicitly couple pointers to our system in analogy to a classical ToF setup.
Based on this setup we will, however, not derive a time-of-flight or time-of-arrival operator directly for the particle.
In fact we extract the momentum operator of the particle from simple observables of the pointers.

In a previous work~\cite{FabioPaper} it was shown that the momentum of a free, non-relativistic quantum particle can be extracted from an operational observable based on the concept of a classical ToF measurement.
This observable has been defined with the help of two continuous systems, serving as pointers in the sense of von Neumann.
In contrast to the projector formalism, it is then not necessary in such an operational approach to decompose the quantum state of the particle in momentum eigenstates and to employ a momentum projector acting on the particle's motional degrees of freedom.
In fact, the momentum information can be read off from the pointers in analogy to the classical ToF concept.

Here, we apply a similar idea but replace the continuous pointers by two-level systems (qubits).
That is, we show that even the most basic quantum systems can be used as measurement devices for the momentum of a particle.
A general concept to apply two-level systems for a measurement of continuous variables has been discussed in an earlier work~\cite{PhysRevA.67.060101}.
In contrast to this work, we study an explicit model and examine which qubit-based observable has to be defined operationally in order to determine the particle's momentum distribution.
Therefore, our model can serve as an addition to quantum measurement theory since it provides another example to the fundamental question on how to gain information from quantum systems.

Moreover, two-level systems are fundamental candidates to study whether their entanglement possibly leads to an improved measurement~\cite{Pedrozo2020}.
These discussions have led to various quantum metrological applications~\cite{PhysRevLett.96.010401,giovannetti2011advances}.
Similarly, we analyze which initial qubit state results in an optimized operational measurement scheme for the particle's momentum.
We show that this state is in fact given by a maximally entangled Bell state.
This novel example demonstrates a further utilization of entanglement in quantum measurement theory.

Our work is organized as follows: In Sect.~\ref{sec:ToFMes} we describe our conceptional setup which will be modeled by a three-body Hamiltonian.
We solve the corresponding dynamics in the Heisenberg picture.
We then discuss in Sect.~\ref{sec:MeasObser} in which way the momentum of the particle can be extracted by measuring certain observables related to the two-level pointers.
In Sect.~\ref{sec:Optim} we contrast and compare the measurement of the particle's momentum distribution with the help of separable and entangled pointers.
In Sect.~\ref{sec:Conclusion} we summarize our work.

\section{Conceptional setup and dynamics}
\label{sec:ToFMes}
\begin{figure}[htp]
    \centering
    \includegraphics[width=\linewidth]{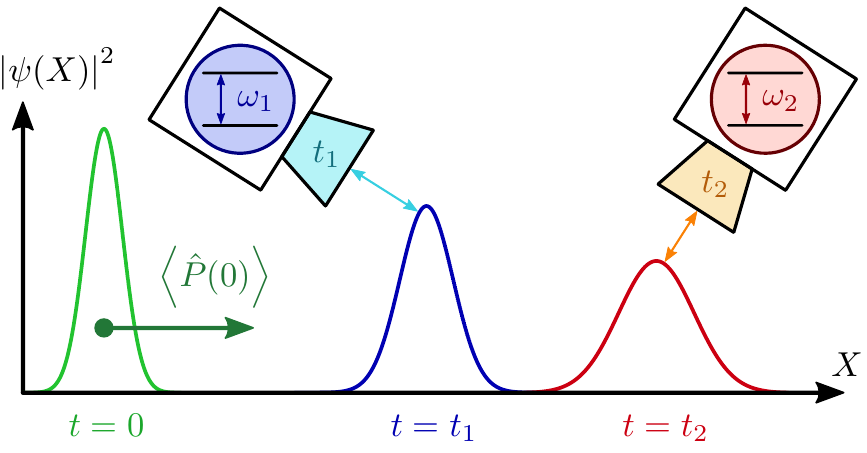}
    \caption{Conceptional setup of a time-of-flight measurement. The quantum particle is visualized in position space for three successive times. 
    Its wave packet $\psi(X)$ is initially prepared at $t=0$ (green wave packet) with an arbitrary momentum distribution. 
    A coupling of the evolved particle (blue and red wave packets) with the two qubit-based pointers is performed at the specific times $t_1$ and $t_2$, with $t_2>t_1>0$. 
    The two-level pointers, which possess transition frequencies $\omega_1$ and $\omega_2$, will record position information of the particle at two different times. This setup is in closest analogy to the classical ingredients to measure the momentum of a massive particle.} 
	\label{fig:SystemKonzept} 
\end{figure}

The concept to measure the momentum of a free quantum  particle based on a pair of two-level systems is shown in Fig.~\ref{fig:SystemKonzept}. 
The quantum mechanical particle represented by its wave packet $\psi(X)$ in position representation is initially prepared at time $t=0$.
It will then be coupled to two qubit-based pointers at the specific times $t_1$ and $t_2>t_1$, respectively. 
Each pointer will record position information of the particle.
We emphasize that a classical ToF measurement can be done similarly to our quantum concept. 
One simply takes two snapshots of the classically moving particle at two different times $t_1$ and~$t_2$. 
The corresponding pictures will show the particle at different positions, and hence, we can determine a mean classical momentum. 
In contrast to the quantum case these two position measurements will not disturb the particle. 
Depending on their specific interaction, a quantum particle will become correlated with the pointer.
Moreover, since we need two pointers for our concept, see Fig.~\ref{fig:SystemKonzept}, they will quantum mechanically influence each other via the particle. 
In order to show how we can exploit these quantum correlations in our setup, we will discuss a suitable particle--pointer interaction in the following paragraph.

\subsection{Particle--pointer interaction}
\label{sec:Interactionsection}
As argued before, an essential part of our concept depicted in Fig.~\ref{fig:SystemKonzept} is the coupling between particle and two-level pointer. 
We will not directly measure the momentum on the free particle. 
Instead we have argued that our two-level pointers need to register position information of the particle's state
\begin{align}
    \kt{\psi}=\int\!\dd X \, \psi(X) \kt{X}
    \label{eq:DefinitionPsivonX}
\end{align}
written in terms of position eigenstates $\kt{X}$ with wave function $\psi(X)$. 
Once position information has been encoded in both pointers, we can read them off using suitable two-level observables.

Let us assume the state
\begin{align}
    \kt{\varphi}_k= a \kt{0}_k+ b \kt{1}_k 
\end{align}
of a single pointer $k\in \left\{1,2\right\}$ to be in a certain superposition of eigenstates 
\begin{align}
    \kt{0}_k=\sigma^z_k \kt{0}_k
    \label{eq:Basiseins}
\end{align}
and
\begin{align}
    \kt{1}_k=-\sigma^z_k \kt{1}_k
    \label{eq:Basiszwei}
\end{align}
of the Pauli-spin operator $\sigma^z_k$. It is then immediately clear that a unitary transformation
\begin{align}
    \begin{split}
        &\ee^{-\ii \kappa \op{X}\otimes \sigma^z_k} \kt{\psi} \otimes \kt{\varphi}_k \\ 
    &= \int\!\dd X \, \psi(X) \left[ a \ee^{-\ii \kappa X} \kt{0}_k+ b \ee^{\ii\kappa X} \kt{1}_k \right] \otimes \kt{X}
    \end{split}
\end{align}
with some coupling constant $\kappa$ transfers position information into the superposition phase of the two-level pointer.
The corresponding interaction Hamiltonian which generates this unitary evolution reads
\begin{align}
    \label{eq:IntXsigma}
	\op{H}_\mathrm{int}= f_k(t) \op{X}\otimes \sigma^z_k 
\end{align}
with a time-dependent coupling function $f_k(t)$ which effectively leads to the constant $\kappa$ and will be specified later for our situation.

We note that bilinear interactions of the form $\op{X} \otimes {\sigma^z}$ are not just suitable from our conceptional point of view but are actually utilized in atom optics, for example in the context of spin coherence~\cite{englert1988spin}.
Indeed, similar interactions have been implemented in order to build an atom interferometric setup~\cite{PhysRevLett.123.083601}.
These specific applications rely on magnetic fields interacting with internal states of the atom and leading to Hamiltonians analogous to Eq.~\eqref{eq:IntXsigma}.

\subsection{Hamiltonian and rescaling}
After having determined the Hamiltonian governing the interaction between particle and pointer, we can write down the complete Hamiltonian by adding 
\begin{align}
\op{H}_\mathrm{free} =\frac{\op{P}^2}{2M}
\end{align} 
for the free motion of the particle with momentum $\op{P}$ and mass~$M$ as well as
\begin{align}
\op{H}_k = \frac{\hbar\omega}{2} \sigma^z_k
\end{align}
representing the free time evolution of the two qubit-based pointers.
Here we assume an identical energy-level spacing $\hbar\omega$ for simplicity.
Note that the following treatment could also be generalized to two different transition frequencies.
To further simplify calculations we introduce dimensionless operators
\begin{align}
	\op{X}^\prime \definer \frac{\op{X}}{a}, \quad \op{P}^\prime \definer  \frac{a}{\hbar} \op{P}, \quad \op{H}^\prime \definer \frac{M a^2 }{\hbar^2} \op{H}
\end{align}
leading to the transformations
\begin{align}
    	t^\prime \definer \frac{\hbar}{Ma^2} t, \quad \omega^\prime \definer \frac{M a^2}{\hbar} \omega, \quad f_k^\prime(t^\prime) \definer \frac{M a^3}{\hbar^2} f_k(t)
\end{align}
in the scalar parameters. The scaling length $a$ in these transformations can in principle be chosen arbitrarily. 
Nevertheless, a natural choice would be the width of initial wave packet~$\psi(X)$, Eq.~\eqref{eq:DefinitionPsivonX}.

In the following calculations, only these rescaled quantities are used and hence we drop the prime from now on for clarity.
The complete Hamiltonian of the system therefore reads
\begin{align}
\op{H}(t)=\frac{\omega}{2} \left(\sigma^z_1+\sigma^z_2 \right) +\frac{\op{P}^2}{2}+\sum_{k=1}^2 f_k(t) \op{X} \otimes \sigma^z_k
\label{eq:Hamiltoniannormiert}
\end{align}
with the free evolution of the three subsystems and the two coupling terms to describe interactions between particle and pointers.

\subsection{Dynamics of the system}
With this Hamiltonian we can now analyze the dynamics of the system. 
The calculation turns out to be most elegant in the Heisenberg picture in which all operators $\op{O}$ will be in general time dependent and obey the equation of motion
\begin{align}
    \frac{\dd \op{O}(t)}{\dd t} = \ii \com{\op{H}(t)}{\op{O}(t)}
\end{align}
given no time dependency of the operator in the Schrö\-ding\-er picture.
Hence, for the operators of our system we arrive at
\begin{align}
\frac{\dd \sigma^z_k (t)}{\dd t} &=0 \label{eq:EquationofMotion1}
\end{align}
and
\begin{align}
\frac{\dd \sigma^\pm_k (t)}{\dd t} &= \pm \ii\omega \sigma_k^\pm (t) \pm {2\ii} f_k(t) \op{X}(t) \otimes \sigma^\pm_k (t) \label{eq:DiffGleichungsigmaplus}
\end{align}
concerning each pointer $k\in\left\{ 1,2\right\}$, where we have introduced the ladder operators
\begin{align}
	\sigma^\pm_k \definer \frac{1}{2} \left( \sigma^x_k\pm \ii \sigma^y_k \right).
	\label{eq:DefSigmaPlusMinus}
\end{align}
The observables of the particle evolve according to
\begin{align}
\frac{\dd \op{P} (t)}{\dd t}  &=-f_1(t) \sigma^z_1(t)-f_2(t) \sigma^z_2(t) \label{eq:EquationofMotion3}
\end{align}
and
\begin{align}
\frac{\dd \op{X} (t)}{\dd t} &= \op{P}(t). \label{eq:EquationofMotion4}
\end{align}
Since the two operators $\sigma_k^z$ commute with the Hamiltonian, they are constants of motion. 
This fact enables us to solve Eqs.~\eqref{eq:EquationofMotion1}, \eqref{eq:EquationofMotion3} and \eqref{eq:EquationofMotion4}.
For a shorter notation we emphasize the operators in the Heisenberg picture with their respective time dependency. 
In contrast, operators without time dependency correspond to the Schrödinger picture and coincide with the initial Heisenberg operators at $t=0$.
The Heisenberg operators can then be written as
\begin{align}
\sigma^z_k(t)&=\sigma^z_k, \label{eq:AllgemeinStart}\\
\op{P}(t)&=\op{P}-a_1(t) \sigma^z_1-a_2(t) \sigma^z_2
\end{align}
and
\begin{align}
\op{X}(t)&=\op{X}+\op{P}t-{b_1(t)} \sigma^z_1-{b_2(t)} \sigma^z_2
\label{eq:AllgemeinEnde}
\end{align}
with coefficients
\begin{align}
a_k(t)&\definer\int\limits_0^t \! f_k(\tau)\,\dd\tau 
\label{eq:CoefficientA}
\intertext{and}
b_k(t)&\definer\int\limits_0^t \! a_k(\tau)\,\dd\tau
\label{eq:CoefficientB}
\end{align}
that depend on the coupling functions $f_k(t)$ of interaction, Eq.~\eqref{eq:Hamiltoniannormiert}.

Linear homogeneous differential equations~\eqref{eq:DiffGleichungsigmaplus} can be integrated to obtain the solutions
\begin{align}
\sigma^\pm_k(t)= \sigma^\pm_k \ex{ \pm \ii\omega t \pm {2\ii} \int\limits_0^t \! f_k(\tau) \op{X}(\tau) \,\dd \tau }\!,
\label{eq:DiracFormelSigmaplusminus}
\end{align}
where the integral in the exponential also depends on the coupling functions.
We now choose delta-like coupling functions
\begin{align}
	f_k(t)=\kappa \delta(t-t_k),
\end{align}
which concentrate the interactions on two specific times $t_k$ and allow us to easily solve the integrations without losing anything fundamental of our concept.
When evaluating the integrals, it is essential to keep track of the time ordering
\begin{align}
	0<t_1<t_2<T
	\label{eq:TimeOrdering}
\end{align}
between the interaction times and the read-off time $T$.
Time-dependent coefficients, Eqs.~\eqref{eq:CoefficientA} and \eqref{eq:CoefficientB}, then read
\begin{align}
a_k(t)&=\kappa \int\limits_0^t \! \delta(\tau-t_k) \,\dd \tau=\kappa \heavi(t-t_k) \label{eq:KoeffizientenEins}\\
\intertext{and}
b_k(t)&=\int\limits_0^t \! a_k(\tau) \,\dd \tau=\kappa (t-t_k)\heavi(t-t_k) \label{eq:KoeffizientenZwei}
\end{align}
for an arbitrary time $t$ written with the help of a \mbox{Heaviside} step function $\heavi(t-t_k)$. 
These coefficients lead to the explicit Heisenberg operators 
\begin{align}
\sigma^z_k(T)&=\sigma^z_k,\\
\op{P}(T)&=\op{P}-\kappa \sigma^z_1 -\kappa \sigma^z_2
\end{align}
and
\begin{align}
\op{X}(T)&=\op{X}+\op{P}T-\kappa(T-t_1) \sigma^z_1 -\kappa(T-t_2) \sigma^z_2
\end{align}
for the read-out time $T$. 
By solving the integrals in Eq.~\eqref{eq:DiracFormelSigmaplusminus} we arrive at
\begin{align}
\sigma^\pm_1(T)=\sigma^\pm_1 \otimes\ee^{\pm \ii \omega T} \ee^{ \pm {2\ii\kappa} \left(\op{X}+{\op{P}}t_1\right) }
\label{eq:Sigmaplusminuseins}
\end{align}
and
\begin{align}
\begin{split}
    \sigma^\pm_2(T)=&\sigma^\pm_2 \otimes\ee^{\pm \ii \omega T} \ee^{ \pm {2\ii\kappa} \left(\op{X}+{\op{P}}t_2\right) }  \otimes\ee^{ \mp {2 \ii \kappa^2}(t_2-t_1) \sigma^z_1 } 
\end{split}
\label{eq:Sigmaplusminuszwei}
\end{align}
in which aforementioned time ordering, Eq.~\eqref{eq:TimeOrdering}, becomes essential.

These Heisenberg operators already contain the initial position and momentum operator of the particle. 
However, we have to combine them to access the momentum operator alone.
Furthermore, this combination has to describe an actual measurement, i.e., it must possess an operational meaning. 
Therefore, we will analyze suitable qubit-based correlation functions in the following section. 

\section{Particle--pointer correlations}
\label{sec:MeasObser}
Using definition, Eq.~\eqref{eq:DefSigmaPlusMinus}, of the ladder operators together with their Heisenberg evolution, Eqs.~\eqref{eq:Sigmaplusminuseins} and \eqref{eq:Sigmaplusminuszwei}, we clearly see that any information on the particle's momentum $\op{P}$ must be contained in a correlation measurement of two-qubit operators
\begin{align}
\begin{split}
    \sigma^{xx}(T) &\equiv \sigma_1^x(T) \sigma_2^x(T) \\ 
    &= \left( \sigma_1^+(T)+\sigma_1^-(T) \right) \left( \sigma_2^+(T)+\sigma_2^-(T) \right) \\ 
    &= \ee^{\ii\op{\mu}} \kb{00}{11} + \ee^{\ii\op{\eta}} \kb{10}{01} + \mathrm{h.c.}
\end{split}
\label{eq:SigmaxxCorrelation}
\end{align}
and
\begin{align}
\begin{split}
    \sigma^{yy}(T) &\equiv \sigma_1^y(T) \sigma_2^y(T) \\ 
    &= - \left( \sigma_1^+(T)-\sigma_1^-(T) \right) \left( \sigma_2^+(T)-\sigma_2^-(T) \right) \\
    &= -\ee^{\ii\op{\mu}} \kb{00}{11} + \ee^{\ii\op{\eta}} \kb{10}{01} + \mathrm{h.c.}
\end{split}
\label{eq:SigmayyCorrelation}
\end{align}
where the exponents are defined via
\begin{align}
    \op{\eta} \definer 2\kappa \op{P} \left(t_2-t_1\right) 
    \label{eq:EtaDef}
\end{align}
and
\begin{align}
    \op{\mu} \definer 2\kappa \left( 2 \op{X}+\op{P} \left(t_1+t_2\right)\right)  + 2 \omega T.
\end{align}
We find that these operators have a very similar structure. 
On the particle they act as unitary rotations governed by operators $\op{\eta}$ and $\op{\mu}$, encoded in two distinct subspaces spanned by the two pairs of two-qubit states \{${\kt{00}\! ,\! \kt{11}}$\} and \{${\kt{01}\! ,\! \kt{10}}$\}.
While the operator $\op{\eta}$ contains the momentum $\op{P}$ in a tunable manner via the time difference $t_2-t_1$, the $\op{\mu}$ operator depends on the position $\op{X}$ as well and is therefore not useful for accessing only the momentum of the particle. 

Hence, a pure momentum measurement can be described by extracting only the $\op{\eta}$ part from these operators. 
In the following section we will show that this extraction can either be achieved by combining several measurements of the two different qubit-based operators or by using initially entangled pointers for a measurement with only one of the two operators. 

\subsection{Characteristic function from qubit correlations}
After having established operationally meaningful two-qubit operators, Eqs.~\eqref{eq:SigmaxxCorrelation} and \eqref{eq:SigmayyCorrelation}, we will now investigate the structure of the corresponding correlation functions and conclude how we can use them to determine the momentum of the particle. 

We assume that particle and pointers are initially not correlated, so that the total state 
\begin{align}
    \rho = \rho_\mathrm{s}\otimes \kb{\varphi}{\varphi}
\label{eq:ProduktStateInitial}
\end{align}
is a product between a mixed state $\rho_\mathrm{s}$ for the particle and the pure two-qubit state 
\begin{align}
    \kt{\varphi} = \varepsilon_{00}\kt{00} + \varepsilon_{01}\kt{01} + \varepsilon_{10}\kt{10} + \varepsilon_{11}\kt{11}
    \label{eq:PointerState}
\end{align}
with arbitrary coefficients $\varepsilon_{ij}$ obeying the normalization condition.
As we want to optimize these pointer states, we assume them to be perfectly preparable.
Using this state we arrive at the correlation functions 
\begin{align}
\begin{split}
    \mean{\sigma^{xx}(T)} = \,\tr{\sigma^{xx}(T)\rho} =\, &\realpart{2 \varepsilon_{01}\varepsilon^*_{10}\trin{\mathrm{s}}{\ee^{\ii\op{\eta}}{\rho_\mathrm{s}}} } \\& + \realpart{2 \varepsilon^*_{00}\varepsilon_{11}\trin{\mathrm{s}}{\ee^{\ii\op{\mu}}{\rho_\mathrm{s}}} }
\end{split}
    \label{eq:Mittelwertsigmaxx}
\end{align}
and 
\begin{align}
\begin{split}
    \mean{\sigma^{yy}(T)} = \,\tr{\sigma^{yy}(T) \rho} = \,&\realpart{2 \varepsilon_{01}\varepsilon^*_{10}\trin{\mathrm{s}}{\ee^{\ii\op{\eta}}{\rho_\mathrm{s}}} } \\ & - \realpart{2 \varepsilon^*_{00}\varepsilon_{11}\trin{\mathrm{s}}{\ee^{\ii\op{\mu}}{\rho_\mathrm{s}}} }
\end{split}
    \label{eq:Mittelwertsigmayy}
\end{align}
of a two-qubit measurement at time $T$ as described by given operators, Eqs.~\eqref{eq:SigmaxxCorrelation} and \eqref{eq:SigmayyCorrelation}. 

When we now recall the definition of $\op{\eta}$, Eq.~\eqref{eq:EtaDef}, we see that the characteristic function
\begin{align}
    C_{\op{P}}(\lambda) = \trin{\mathrm{s}}{\ee^{\ii\lambda \op{P}}\rho_\mathrm{s}}
    \label{eq:CharacteristicFunction}
\end{align}
of the particle's momentum $\op{P}$ occurs in both correlation functions with a parameter $\lambda \equiv 2\kappa (t_2-t_1) > 0$, which can be tuned by varying the interaction times $t_k$. With the properties $C_{\op{P}}(-\lambda) = C^*_{\op{P}}(\lambda)$ and $C_{\op{P}}(0) = 1$ of a characteristic function we can calculate its Fourier transform which results in the momentum distribution
\begin{align}
    \trin{\mathrm{s}}{\rho_\mathrm{s} \kb{P}{P}} = \br{P}{\rho_\mathrm{s}}\kt{P}= \frac{1}{2\pi} \int_\R\! C_{\op{P}}(\lambda)\ee^{-\ii\lambda P}\dd \lambda
\end{align}
of the incoming particle.
Hence we clearly see that our two-qubit correlation functions, Eqs.~\eqref{eq:Mittelwertsigmaxx} and \eqref{eq:Mittelwertsigmayy}, deliver full momentum information of the particle.
At this point we emphasize once more that our approach needs no definition of a time-of-flight operator.
Rather we have used a setup with the idea of a classical ToF experiment to make the exact momentum distribution of the particle accessible through ensemble measurements on qubit pointers.

In the next section it remains to be shown how characteristic function, Eq.~\eqref{eq:CharacteristicFunction}, has to be extracted finally from correlation functions based on suitably prepared qubit pointers, i.e., by choosing suitable coefficients $\varepsilon_{ij}$ for the two-qubit state $\ket{\varphi}$, Eq.~\eqref{eq:PointerState}. 

\section{Pointer states for momentum measurement}
\label{sec:Optim}
Correlation functions, Eqs.~\eqref{eq:Mittelwertsigmaxx} and \eqref{eq:Mittelwertsigmayy}, are clearly operational quantities. 
However, in order to isolate characteristic function, Eq.~\eqref{eq:CharacteristicFunction}, a single correlation function will not be sufficient.
In this section we will demonstrate two versions to operationally reconstruct the particle's characteristic function from qubit-based correlations.
In fact, we need four correlation measurements if we choose separable pointers and just two if we prepare them in an entangled state.

\subsection{Separable pointers}
We begin by investigating the case of initially separable qubit states $\kt{\varphi}$, Eq.~\eqref{eq:PointerState}.
That is, we prepare the two measurement devices independently.
The contribution of the characteristic function will be optimized by increasing the amplitude $\abs{\varepsilon_{01}\varepsilon_{10}}$ in both correlation functions, Eqs.~\eqref{eq:Mittelwertsigmaxx} and \eqref{eq:Mittelwertsigmayy}. 
In the case of separability the condition of normalization requires  $\abs{\varepsilon_{01}}+\abs{\varepsilon_{10}}\leq 1$.
Hence, the maximum will be reached for $\abs{\varepsilon_{01}}=\abs{\varepsilon_{10}}={1}/{2}$, which yields the optimized separable pointer state
\begin{align}
    \kt{\varphi}_\mathrm{sep} = \frac{1}{2}\left( \kt{0} + \ee^{\ii\phi_1} \kt{1}\right)\otimes\left(\kt{0} + \ee^{\ii\phi_2} \kt{1} \right)
    \label{eq:OptimisedSepState}
\end{align}
with the relative phases $\phi_k$ of the single qubits.
For this choice we can sum up correlation functions, Eqs.~\eqref{eq:Mittelwertsigmaxx} and \eqref{eq:Mittelwertsigmayy}, and arrive at 
\begin{align}
    \mean{\sigma^{xx}(T)}_\mathrm{sep} + \mean{\sigma^{yy}(T)}_\mathrm{sep} = \realpart{C_{\op{P}}(\lambda) \ee^{\ii(\phi_2-\phi_1)}}.
    \label{eq:MeanSep}
\end{align} 
Eventually, by tuning the relative phase of the two single-qubit states it is possible to measure the full characteristic function in the sense of its real and imaginary part. 
Choosing $\phi_1-\phi_2 = 0$ yields the real part $\realpart{C_{\op{P}}(\lambda)}$, while the phase $\phi_1-\phi_2 = \frac{\pi}{2}$ results in a measurement of the imaginary part $\imgpart{C_{\op{P}}(\lambda)}$.
We emphasize that this measurement scheme requires four ensemble measurements in total: two correlation functions and each for two settings of the relative phases.

Moreover, the time difference $t_2-t_1$ of interaction times needs to be tuned in order to reconstruct the characteristic function $C_{\op{P}}(\lambda)$ in the parameter regime $\lambda>0$.

This result raises the question of whether dropping the condition of separability for the pointers will lead to an advanced measurement procedure.

\subsection{Entangled pointers}
Dropping the condition of separability means that the qubits are prepared in an initially entangled way.
Just as before, we construct the state in such a way that it maximizes the amplitude $\abs{\varepsilon_{01}\varepsilon_{10}}$. 
The normalization condition in the case of arbitrary states $\kt{\varphi}$, Eq.~\eqref{eq:PointerState}, is less restrictive than in the case of separability, and the maximum is reached for \linebreak $\abs{\varepsilon_{01}}=\abs{\varepsilon_{10}}={1}/{\sqrt{2}}$.
Hence, the optimal pointer state now reads
\begin{align}
    \kt{\varphi}_\mathrm{ent}=\frac{\kt{01}+\ee^{\ii\phi}\kt{10}}{\sqrt{2}},
    \label{eq:OptimierterStateEnt}
\end{align}
which is a maximally entangled state with relative phase~$\phi$. 
Inserting this state in correlation function, Eq.~\eqref{eq:Mittelwertsigmaxx}, yields
\begin{align}
    \mean{\sigma^{xx}(T)}_\mathrm{ent}  = \realpart{C_{\op{P}}(\lambda) \ee^{-\ii\phi}}
    \label{eq:MeanEntxx}
\end{align}
as well as an analogous expression
\begin{align}
    \mean{\sigma^{yy}(T)}_\mathrm{ent}  = \realpart{C_{\op{P}}(\lambda) \ee^{-\ii\phi}}.
    \label{eq:MeanEntyy}
\end{align}
With entangled pointers the characteristic function can be reconstructed from just a single correlation function.

The real and imaginary parts of the characteristic function are obtained as before by taking $\phi=0$ and $\phi=\frac{\pi}{2}$.
Hence, in total we now need just two ensemble measurements for each parameter $\lambda > 0$.

\section{Conclusion}
\label{sec:Conclusion}
The main result of the present work is a qubit-based point\-er model, which represents a von Neumann approach to measure the momentum of a massive particle.
The model is based on the classical version of a time-of-flight concept.
We have explicitly solved its quantum dynamics in the Heisenberg picture. This solution has allowed us to identify suitable qubit-based observables which contain information on the complete momentum distribution of the measured particle via its characteristic function.
Moreover, we have shown that this information can be extracted more efficiently if we prepare the two-level pointers in a Bell state. 

Our study provides an analytically solvable model for an early idea of von Neumann, while at the same time presenting a connection to entanglement.
An interesting extension would be the inclusion of external potentials acting on the measured particle.
One could then specify operational observables for freely falling or trapped particles.
Moreover, a three-dimensional generalization would possibly allow us to sketch a concept of an operational angular momentum measurement.

\begin{acknowledgements}
	Fabio Di Pumpo acknowledges the financial support of the project ``Metrology with interfering Unruh-DeWitt detectors'' (MIUnD) which is funded by the Carl Zeiss Foundation (Carl-Zeiss-Stiftung), the IQ\textsuperscript{ST} which is financially supported by the Ministry of Science, Research and Art Baden-W\"urttemberg (Ministerium f\"ur Wissenschaft, Forschung und Kunst Baden-W\"urttemberg) and the QUANTUS project which is supported by the German Aerospace Center (Deutsches Zentrum f\"ur Luft- und Raumfahrt, DLR) with funds provided by the Federal Ministry of Economic Affairs and Energy (Bundesministerium f\"ur Wirt\-schaft und Energie, BMWi) due to an enactment of the German Bundestag under grant no. 50WM1956 (QUANTUS V).
\end{acknowledgements}

\section*{Author contributions}
All authors were involved in the preparation of the manuscript, and all authors have read and approved the final manuscript.

\section*{Data Availability Statement}
This manuscript has no associated data, or the data will not be deposited. [Authors' comment: All calculations have been performed analytically with the help of the presented equations.]

\bibliography{Literatur.bib}

\end{document}